Cross-phenotype meta-analysis reveals large-scale *trans*-eQTLs mediating patterns of transcriptional co-regulation
Short: *trans*-eQTLs reveal patterns of transcriptional co-regulation

Boel Brynedal[1,2†], Towfique Raj[2,3,4,5], Barbara E Stranger[2,3,5‡], Robert Bjornson[6], Benjamin M Neale[2,5,7], Benjamin F Voight[8], Chris Cotsapas[1,2,9]*

1 Department of Neurology, Yale University School of Medicine, New Haven CT 06520
2 Program in Medical and Population Genetics, Broad Institute, Cambridge, MA 02142
3 Division of Genetics, Brigham and Women's Hospital, Boston, MA 02115
4 Department of Neurology, Brigham and Women's Hospital, Boston, MA 02115
5 Harvard Medical School, Boston, MA 02115
6 Department of Computer Science, Yale University, New Haven CT 06510
7 Analytical and Translational Genetics Unit, Massachusetts General Hospital, Boston, MA 02114
8 Department of Pharmacology and Department of Genetics, Perelman School of Medicine, University of Pennsylvania, Philadelphia PA 19104
9 Department of Genetics, Yale University School of Medicine, New Haven CT 06520

† present address: Karolinska Institutet, Stockholm, Sweden
‡ present address: Section of Genetic Medicine, Department of Medicine, University of Chicago, Chicago IL 60637

* Correspondence to cotsapas@broadinstitute.org

First paragraph word count: 131
Rest of main manuscript word count:
Online methods:
Supplementary files:
Tables (with legend): 2
Figures (with legend): 2

**Genetic variation affecting gene regulation is a central driver of phenotypic differences between individuals and can be used to uncover how biological processes are organized in a cell. Although detecting *cis*-eQTLs is now routine, *trans*-eQTLs have proven more challenging to find due to the modest variance explained and the multiple tests burden of testing millions of SNPs for association to thousands of transcripts. Here, we successfully map *trans*-eQTLs with the complementary approach of looking for SNPs associated to the expression of multiple genes simultaneously. We find 732 *trans*-eQTLs that replicate across two continental populations; each *trans*-eQTL controls large groups of target transcripts (regulons), which are part of interacting networks controlled by transcription factors. We are thus able to uncover co-regulated gene sets and begin describing the cell circuitry of gene regulation.**

Biological processes are carefully orchestrated events requiring precise activation and repression of participating genes by hierarchical gene regulation mechanisms. This elaborate co-regulation can be seen in the complex patterns of gene co-expression across tissues[1] and conditions[2]; the overlap and organization of transcription factor target sets[3]; and the organization of gene interaction networks[4]. Furthermore, it has become apparent that a substantial fraction of common genetic variants driving organismal traits such as disease risk affect gene regulatory sequences rather than coding sequence[5,6]. Thus understanding how genetic variation influences the co-regulation of multiple genes will reveal the major regulators of biological processes and the molecular mechanisms underlying organismal traits, including disease susceptibility.

Expression quantitative trait locus (eQTL) mapping[7,8] is now routinely used to identify variants controlling gene expression *in cis* (*cis*-eQTLs), which are thought to perturb regulatory sequences. In contrast, variation affecting the regulatory machinery rather than regulatory sequences around a gene should map elsewhere in the genome. Whilst a handful of such *trans*-eQTLs have been described in humans[9-12], they do not account for the fraction of transcript level heritability attributed to *trans*-acting variation[13]. This is likely because, unlike *cis*- eQTLs where analysis can be limited to the genomic locus encoding each gene, *trans*-eQTL discovery requires genome-wide testing, and the correction required for testing millions of SNPs for association to thousands of transcripts renders even large studies underpowered. Thus, simple analyses which look for independent association evidence that many transcripts map to the same region will fail, and new approaches are required to sensitively detect *trans*-eQTLs and describe patterns of gene co-regulation.

Here, we hypothesize that if variation *in trans* acts on the transcriptional regulation machinery, it should affect multiple transcripts simultaneously and thus allow us to detect co-regulated gene sets, or regulons. We pursue this idea by applying a second-level significance testing[14] framework to detect *trans*-eQTLs affecting multiple transcripts[15]: in essence, our approach is to look for a shift in the distribution of eQTL association statistics at each SNP, indicative of association to multiple transcripts. We are thus not limited either by power to detect association to individual transcripts at genome-wide significance levels nor do we have to test all possible combinations of transcripts for association as in a standard meta-analysis[16]. In extant eQTL data we are able to identify 732 *trans*-eQTLs that replicate across two continental populations and that modulate the expression levels of multiple genes. We further show that these regulons have independent evidence of co-regulation and co-ordinate function on several levels. We are thus able to uncover sets of co-regulated genes in studies of realistic size and begin describing the large-scale organization of the transcriptome.

## Results
### Trans-eQTLs are common and affect hundreds of genes
Our premise is that if a SNP affects the transcript levels of multiple genes, then there should be an excess of low association statistics at that SNP[14]. At each SNP, we test a single hypothesis: whether the association *p*-values deviate from the expected distribution, which we define empirically to account for the correlation between gene expression levels (see Online Methods). We apply this approach to eQTL data from the CEU and YRI HapMap populations (109 and 108 individuals respectively[17]), replicating observations across the two datasets. We consider 8,368 transcripts detected in lymphoblasoid cell lines and 610,180 autosomal SNPs with minor allele frequency >15% in both populations. After data quality control and normalization, we calculate SNP-expression association statistics accounting for gender, and for each SNP calculate a CPMA statistic from the 8368 eQTL *p*-values. As CPMA is a distributional test and does not identify which transcripts are being affected, we identify *trans*-eQTL targets in a separate step using a two-group mixture model on the *p*-value distribution. We define genes as targets if they have >80% probability of belonging to the non-null group (see Online Methods).

We found 26792/610180 SNPs in CEU and 28013/610180 in YRI have an empirical $P_{CPMA}$ < 0.05, with a significant overlap of 1311 SNPs between these sets (hypergeometric *p* = 0.0079). After accounting for linkage disequilibrium ($r^2$ < 0.2), we find that these represent 732 independent effects present in both populations (Figure 1A, B, Table S1), each affecting many genes (Figure 1C). If these effects are genuine *trans*-acting eQTLs, we expect them to fulfill two key predictions across the two populations: the genes they influence should be the same in the two populations; and the direction of effect of the minor allele should be consistent between the two populations for these genes. To account for the correlation between expression levels, we have developed empirical approaches to assess both these predictions (see Online Methods).

For the first prediction, we assess whether the *trans*-eQTL has same targets across the two populations. Given $N_{CEU}$ and $N_{YRI}$ target genes, we observe an intersect $N_{overlap}$ = $N_{CEU}$ {union symbol} $N_{YRI}$. We construct an expected distribution of $N_{overlap}$ across all 610,180 SNPs in the analysis and compute the empirical *p*-value $P_{overlap}$ from this, which accounts for correlations between markers and between expression traits. We find that 329/732 effects have significant overlap of target genes in the two populations (SNP-wise $P_{overlap}$ < 0.05; binomial probability of this number of overlaps occurring by chance $p$ = 1.6 x $10^{-200}$), showing that we can replicate our *trans*-eQTLs across two populations. For the second, we assess whether the direction of effect is consistent across the two populations and find that 618/732 SNPs have SNP-wise *p* < 0.05 by this measure (binomial probability of this number of overlaps occurring by chance $p$ < 1 x $10^{-300}$). We thus show that our approach detects genuine *trans*-eQTLs active in two continental populations and each affecting regulons comprised of hundreds of genes.

### *trans*-eQTLs act via *cis*-eQTLs

We next hypothesized that *trans*-acting variants would exert their effects on their targets by perturbing genes *in cis*[11]. We found that they are located closer to genes than expected by chance ($p = 0.016$), and are more likely to be found within gene boundaries ($p = 0.011$). Overall, we found that 486/732 *trans*-acting variants are within 500 kb of a gene and that we had classified a proximal gene as a regulon member for 154 of these 486, suggesting that a sizeable fraction of our *trans*-acting variants act as *cis*-eQTLs. We found that *trans*-eQTLs were more likely to affect the proximal gene compared to other SNPs with similar CPMA statistics (enrichment of true *trans*-acting SNPs as *cis*-eQTLs $p = 7.2 \times 10^{-8}$ and $p = 2.1 \times 10^{-270}$ in CEU and YRI respectively; Figure 2A). However, *cis*-eQTLs occur frequently and we wanted to establish whether our *trans*-acting SNPs were true *cis*-eQTLs or whether they merely reside in regions enriched for *cis*-eQTL activity. We therefore tested if they had stronger association to their *cis*-targets than surrounding markers and found this to be overwhelmingly the case (rank test $p = 4.3 \times 10^{-43}$ and $p = 5.6 \times 10^{-57}$ in CEU and YRI respectively; Figure 2B), showing they are likely to be the causal alleles for the observed *cis*-eQTL. We thus show that a substantial proportion of the detected *trans*-acting SNPs act as *cis*-eQTLs on proximal genes; this may be the mechanism by which they effect the expression of target genes encoded elsewhere in the genome.

### *trans*-eQTL targets are co-regulated

By definition, our results imply that the genes in our 732 regulons are coordinately regulated. If so, regulon members should share transcriptional control machinery, so we hypothesized that we could detect co-regulation as an enrichment of transcription factors binding to members of each regulon. We test this by looking for enrichment of transcription factor binding events near target genes in lymphoblastoid cell lines profiled by the ENCODE project. We find that 681/732 (93%) target sets are enriched for targets of at least one transcription factor, and 563/732 (77%) show enrichment of two or more transcription factors (Table S2). Of course, these results do not mean the enriched transcription factors induce the *trans*-eQTLs directly, but that the target genes are co-regulated.

We further hypothesized that regulons represent functional gene modules and sought evidence of functional interaction between members of each set. We looked for significant protein-protein interactions between regulon members using DAPPLE[18] and found that 101/732 (14%, binomial $p = 8.1 \times 10^{-20}$) showed significant ($p < 0.05$) increase in direct connections between proteins, constituting a strong enrichment (Figure S1A). Furthermore, 155/732 (21%, binomial $p = 1.8 \times 10^{-52}$) have evidence of participating in broader networks (DAPPLE's indirect networks, Figure S1B). These results support the model that *trans*-eQTLs perturb functionally linked gene sets by altering their regulation at the transcriptional level.

Next, we define a subset representing networks of interacting genes controlled by the same transcriptional machinery. The 36 regulons shown in Table 1 have significant target overlap, effect directionality, target protein connectivity and transcription factor binding in both CEU and YRI. We see a strong correlation between functional interaction at the protein level and transcriptional regulation among target genes: in 35/36 of these regulons, the genes forming direct DAPPLE protein-protein interaction networks are also enriched for targets of significant ENCODE transcription factors. We find that these target gene sets are in general enriched for Gene Ontology terms associated with cellular homeostasis, suggesting that *trans*-eQTLs may often affect basic cell processes such as mRNA and DNA homeostasis during transcription and cell division (Table S3). Thus, these results show that *trans*-acting eQTLs modulate transcriptionally coherent groups of genes involved in basic cellular processes.

Finally, we tried to uncover what molecular changes *trans*-acting variants induce to influence their targets. Concentrating on the 36 regulons described above, we reasoned that the *trans*-acting variants alter the expression of the ENCODE transcription factors enriched for binding to each regulon. We hypothesized that the expression levels of each enriched factor should correlate to regulon genes. We found at least one significant correlation for 32/36 sets (Table 2). We found no evidence that the *trans*-acting variant associates with expression of these transcription factors directly, consistent with our above observation that these variants act on other genes *in cis* to exert their effects (Table 3). We see similar results when considering all 732 regulons (table S2). we therefore suggest that this perturbation eventually results in alteration to transcription factor expression levels and hence changes target transcript expression.

## Discussion

In this work we are able to identify *trans*-eQTLs by looking for evidence that a SNP simultaneously influences many transcripts. We are able to show compelling evidence for widespread *trans*-acting eQTLs in human cells replicating across continental populations, by addressing the massive multiple testing burden inherent to eQTL analyses. We find that large-scale co-regulation of gene sets is a common feature of the human transcriptome[3,19], which can be uncovered by genetic analysis across populations. Some regulons vary across the population in concert with changes to transcription factor expression levels, although other control mechanisms likely exist.

*trans*-eQTLs have proven challenging to detect in human data, despite the substantial heritability of gene expression attributed to them[13]. The modest effect sizes of *trans*-acting variants[10,11] in part drive this failure, as does the systematic noise in gene expression assays[20]. Whilst both issues can be addressed by increasing sample size to boost statistical power[12], the cost and logistics of ascertaining large cohorts makes this approach economically prohibitive, especially when considering multiple tissues[21], so new analytical approaches are required. When considering multiple phenotypes simultaneously, as is the case with *trans*-eQTLs controlling many transcripts, additional information from the expectation of shared association to boost power[16]. Our approach leverages this additional information and we are able to show – in publicly available data in which standard analyses have not yielded results – that *trans*-eQTLs are detectable in large numbers.

Our results are consistent with precise regulation of biological processes, particularly basic homeostatic mechanisms, which may have high-level phenotypic consequences. These observations further support the notion that regulation of basic cell processes is highly orchestrated and occurs on several levels simultaneously[22]. Applying this approach to eQTL datasets from diverse tissues will yield rich insights into tissue-specific regulatory circuits driving diverse cellular processes. Finally, we note that biological exploration and dissection of these pathways will require new experimental tools, which can address the subtleties of quantitative regulatory changes in large numbers of genes.

**Author contributions**
Designed analyses: BB and CC
Performed analyses: BB
Contributed methods: BMN and BFV
Contributed technical expertise: RB
Contributed data: TR and BES
Wrote the paper: BB and CC with contributions by the other authors

**Acknowledgements**
Computing resources at Yale were funded partly by NIH grants RR19895 and RR029676-01.

**Figure legends**

**Figure 1**. 732 *trans*-eQTLs replicate across two continental HapMap populations and affect large regulons. A: Empirical CPMA statistics in CEU and YRI populations. Lines indicate *α < 0.05*. B: Observed number of genes in 732 significant, independent *trans*-eQTL regulons. *trans*-acting variants target the same transcripts across populations (panel C: empirical *P*-value distribution for target overlap between the two populations for 732 *trans*-eQTLs), and have the same allelic effect on their targets (panel D: empirical *P*-value distribution for effect direction consistency in the two populations).

**Figure 2**. *trans*-eQTL SNPs are also *cis*-eQTLs. To understand how *trans*-acting SNPs may exert their effects on distal targets, we asked if they had evidence of altering the expression of genes *in cis*. We found that they had more significant cis-eQTL p-values compared to non-replicating SNPs with similar CPMA scores in each population (panels A and B; rank *p*-values compared to similar CPMA SNPs for CEU and YRI respectively). We also found that they were amongst the strongest *cis*-eQTL signals in their loci (panels C and D: rank-based *p*-values for magnitude of cis-eQTL test statistics compared to all other SNPs in the *trans*-acting locus).

**Table legends**

**Table 1**: Across-population transcriptionally coherent *trans*-eQTLs affect interacting gene sets bound by transcription factors. This narrow subset of 36 regulons satisfies all our predictions: *trans*-eQTL effects are significant in both CEU and YRI (columns 4 and 5); regulons are made up of the same genes in the two populations (columns 6-10) and the allelic effects are consistent in direction (column 11); and target genes form significant interaction networks (assessed by DAPPLE[18]; column 12). Each target set is also enriched for binding of at least one transcription factor in ENCODE chromatin immunoprecipitation data (column 13, further detailed in Table 2). We note that other trans-eQTLs, which do not fulfill the interaction network or transcription factor enrichment criteria, likely modulate their targets through other mechanisms. The explained variance in regulon member expression levels is shown in columns 14-17.

**Table 2**: transcription factor expression levels correlate to regulon genes in 32/36 transcriptionally coherent regulons across populations. For each factor enriched for binding near regulon member genes, we correlated expression level in CEU and YRI to the centroid of regulon expression values (calculated as the first principal component; see Online Methods for details). For the majority of comparisons the correlation was significant (empirical $P < 0.05$; ns = not significant) and only 4/36 transcriptionally coherent regulons did not have at least one transcription factor correlated to regulon member expression. Thus, changes in expression to regulon members appear mediated by changes to expression of relevant transcription factors.

**Table 3**: likely mediators of transcriptionally coherent *trans*-eQTLs. In 15/36 *trans*-eQTL loci, we found evidence that a gene within 500kb of the *trans*-acting SNP could mediate the downstream effect on regulons. Genes were *cis*-eQTLs associated to the trans-acting SNP; were part of the regulon; were significantly correlated to the regulon's overall expression pattern across individuals; or were significant seeds in the DAPPLE[18] interaction networks of the regulon members. In six instances, multiple genes are potential mediators of the *trans*-eQTL effect; this may reflect a broader disruption to the local regulatory landscape induced by the *trans*-acting SNP.

Supplementary materials

Figure S1: Connectivity between regulon members. Plots show histograms of DAPPLE connectivity *p*-values for (A) direct protein connectivity and (B) indirect protein connectivity for 732 significant regulons.

Table S1: Significant *trans*-eQTLs replicating across two populations. Empirical CPMA *p*-values for CEU and YRI are listed in columns 4 and 5. Number of regulon members in CEU and YRI are listed in columns 6 and 7, and the observed overlap in column 8. The expected random number of overlapping regulon members including the 95% confidence interval are included in column 9. The *p*-value of *trans*-eQTL target overlap between CEU and YRI is listed in column 10. Column 11 lists the empirical *p*-value for allelic effect consistency across populations, and column 12 the p-value for protein-protein interactions (assessed by DAPPLE). Column 13 lists the number of transcription factors whose targets are enriched among regulon members (further details in Table S2).

Table S2: ENCODE transcription factor binding enrichment in *trans*-eQTL regulons. For each transcription factor enriched for binding near regulon member genes (number of genes in column 3, enrichment *p*-value in column 4), we correlated its expression level in CEU and YRI to the first principal component of regulon gene expression. The significant correlation coefficients for CEU and YRI are listed in column 5 and 6 ('ns': not significant, 'NA' indicates TF was not detected in the gene expression data set).

Table S3: Gene ontology (GO) terms enriched in more than 20 of the 36 high confidence trans-eQTL regulons. We list the GO ID, GO term and number of regulons for which the term is enriched.

*Online Methods, Brynedal et al.*

*Data processing*
Genotype and gene expression data are described in Stranger et al. 2012 and are publicly available at Array Express (http://www.ebi.ac.uk/arrayexpress/) under accession numbers E-MTAB-198 and E-MTAB-264. This dataset comprises whole genome gene expression quantified from lymphoblastoid cell lines of individuals of the HapMap populations{InternationalHapMapConsortium:2005cu} (Coriell, Camden, New Jersey, United States). We used data from 109 Caucasians living in Utah USA, of northern and western European ancestry (CEU), and 108 Yoruba in Ibadan, Nigeria (YRI). Genotype data was filtered based on minor allele frequency of 15% in both CEU and YRI population, resulting in 610180 SNPs genome wide. Gene expression data was normalized using the GCMA algorithm and filtered based on inter quartile range (iqr) and mean expression values to exclude non-expressed genes and unvarying genes. The IQR was clearly bimodal, and a 0.25 cut of was used (Figure S1.A). The mean expression level of genes was also bimodal, and an intensity cut off of 8 was employed.

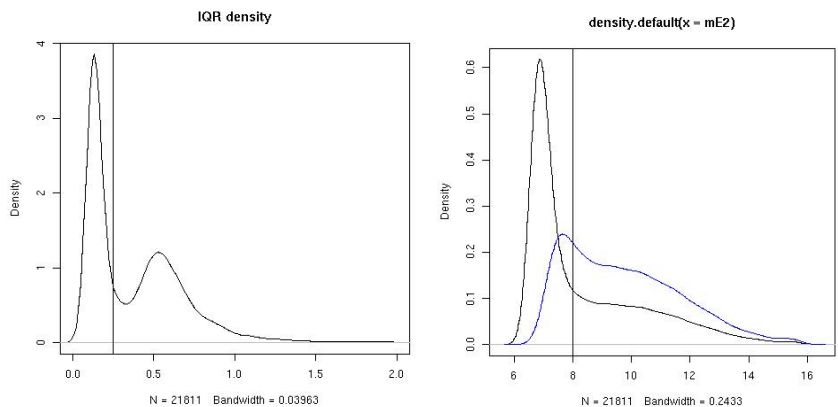

Figure SMR1. Gene expression filtering. A. inter quantile range density distribution where the vertical line indicates the cut of used. B. Density distribution of the full set of genes (prior to IQR filtering in black and post in blue). Vertical line indicates the cut off employed.

*CPMA calculation (MAF correction, CPMA simulation).*
In order to detect SNPs associated to the expression of multiple genes we employed a cross phenotype meta analysis{Cotsapas:2011hb} based on the eQTL p-values from a linear regression with sex as covariate. Calculated CPMA values were corrected by minor allele frequency of the analyzed SNP.

The small sample size and large extent of correlation between genes causes deviance from the expected null-distribution of uniformity. We therefore created a simulated CPMA data set based on the observed covariance of Z scores between genes. First, 8368 correlated Z score vectors were created based on the Cholesky decomposition of the covariance matrix. These were transformed to *p*-values, and CPMA calculated.

Observed CPMA values above 95 percentile of the simulated values were considered significant in each population. We thereafter focused on the set of SNPs that showed significant CPMA statistics in both the CEU and YRI data sets (significant overlap, hypergeometric p-value 0.0079). In order to focus on independent effects we grouped variants based on linkage disequilibrium ($r^2 < 0.2$) and CPMA p-value using the clumping procedure in **plink**{PURCELL:2007dg}. This resulted in 732 independent *trans*-eQTLs and a background set of 83,843 SNPs.

*Defining regulon members and significance testing across populations*
Our CPMA statistic identifies SNPs associated to many transcripts, but not the transcripts themselves. To identify regulon members, for each *trans*-acting SNP we model the *p*-value distribution as a mixture of two

Gaussian distributions with unequal mode and variance using the mclust package in R{Team:2005wf}. We then classified genes as belonging to one of these two distributions, requiring >80% probability to be included in the distribution with low mean *p*-values, and defined genes in this group as *trans*-acting SNP targets. We then defined a *trans*-eQTL regulon as the transcripts thus classified in both CEU and YRI (the intersect set).

We expect true *trans*-eQTLs to affect the same genes across populations. To test the significance of observed overlaps between populations, however, we must account for the fact that transcript expression levels are correlated across individuals, which violates the assumptions of most overlap tests. We therefore assessed the overlap in gene targets of each *trans*-eQTL empirically while taking the observed correlation into account. Given $N_{CEU}$ and $N_{YRI}$ target genes, we recorded the overlap of genes between the top $N_{CEU}$ and $N_{YRI}$ genes at each SNP across the genome, and the significance of the observed overlap was assessed given this null distribution. A histogram of empirical p-values is shown in Figure 1C, in total 329 of the 732 regulons had a p-value < 0.05. This null distribution of overlaps was used to estimate the expected random regulon size as included in Table 1 and S1, column 9. We also estimate the proportion of variance explained as the adjusted $R^2$ from the linear regression models.

Due to the correlation between expression levels, we also empirically test our prediction of allelic effect direction being the same across populations. For each regulon, we calculate the proportion of same-direction effects (using the sign of the linear regression coefficient in our eQTL calculation). We compare this to a null expectation calculated as the proportion of same-direction effects for the same genes at each SNP in the genome.

We tested enrichment of Gene Ontology (GO) categories using conditional hypergeometrical tests as implemented in the R package GOstats [[S Falcon and R Gentleman. Using GOstats to test gene lists for GO term association. Bioinformatics, 23(2):257-8, 2007.]] We focused this analysis on the genes whose proteins were significantly interacting according to the dapple analysis (p-value < 0.05) in order to detect the functionality of the core regulon genes.

*Detecting protein-protein interactions between regulon members with DAPPLE*
We ran DAPPLE{Rossin:2011gq} for each target set of the significant *trans*-eQTLs under default settings. Enrichment of low *p*-values was calculated using CPMA and showed significance for both direct connections and indirect degree (p-values $4.8 \times 10^{-21}$ and $3.6 \times 10^{-70}$ respectively).

*Transcription factor (TF) target overlap*
We used publicly available ENCODE data{Gerstein:2012fq}, where each TF has been connected to target genes targets are defined using the probabilistic model TIP{Cheng:2011bd}. LCL data were downloaded from http://encodenets.gersteinlab.org/enets8.GM_proximal_filtered_network.txt.
For each of the 732 *trans*-eQTL target sets, we investigated the overlap with each of the 50 ENCODE-derived TF target sets. To estimate an empirical p-value for this overlap we created an null distribution for each regulon by randomly selecting genes with similar expression levels and obtaining the random overlap with the TF target set. The observed overlap was then compared to this distribution. We next determined whether enriched TFs were correlated to the expression levels of regulon members (first principal component of the scaled expression levels). Significant Pearson correlations are included in Table S2. Finally, we used a hypergeometric test to detect whether regulon members forming DAPPLE networks are targets of ENCODE TFs.

*cis-eQTL enrichment*
BEDtool [[ref]] was used to collect the closest distance between assayed SNPs and genes (GENCODE v.18) as well as the overlap between genes and SNPs. The distance to closest gene was not correlated to the minor allele frequency of the SNPs. We therefore compared the distance for the 732 *trans*-eQTLs to the full set of independent SNPs using Wilcoxon non-parametric rank tests. Number of actual overlaps with features was compared using binomial tests.

We tested the enrichment of *cis*-eQTL effects among our *trans*-eQTLs in two complimentary ways. First, we assessed whether *trans*-eQTLs were more likely to affect a nearby gene compared to other SNPs with similar CPMA scores. For each *trans*-acting SNP, we compared eQTL *p*-values for genes within 500 kb; for each gene we compared the strength of association to 1000 randomly selected SNPs with similar CPMA scores to the *trans*-acting SNP.

Secondly, we tested whether the *trans*-eQTL was the strongest *cis*-eQTL effect in the region. We therefore compared the strength of association between all SNPs within 500 Mb of the suspected *cis*-gene to the strength of association of the *trans*-eQTL..P-values were calculated as the rank of the *trans*-eQTLs *p*-value divided by the number of regional SNPs.

Putative *trans*-eQTL mediators in *cis*: In order to detect additional potential *cis*-eQTL mechanisms underlying the *trans*-eQTLs we looked into the genes located within 500 kb of a set of 36 *trans*-eQTLs with significant across-population target overlap, significant across-population target directionality, significant TF target overlap, and significant dapple direct connectivity scores. We calculated the correlation between their expression levels and the first principal component of the expression levels (scaled) of the regulon members. We derived an empirical p-value for the correlation between potential 'cis-genes' and regulon members by permuting sample labels for the *cis*-gene and recalculating the correlation 1000 times.

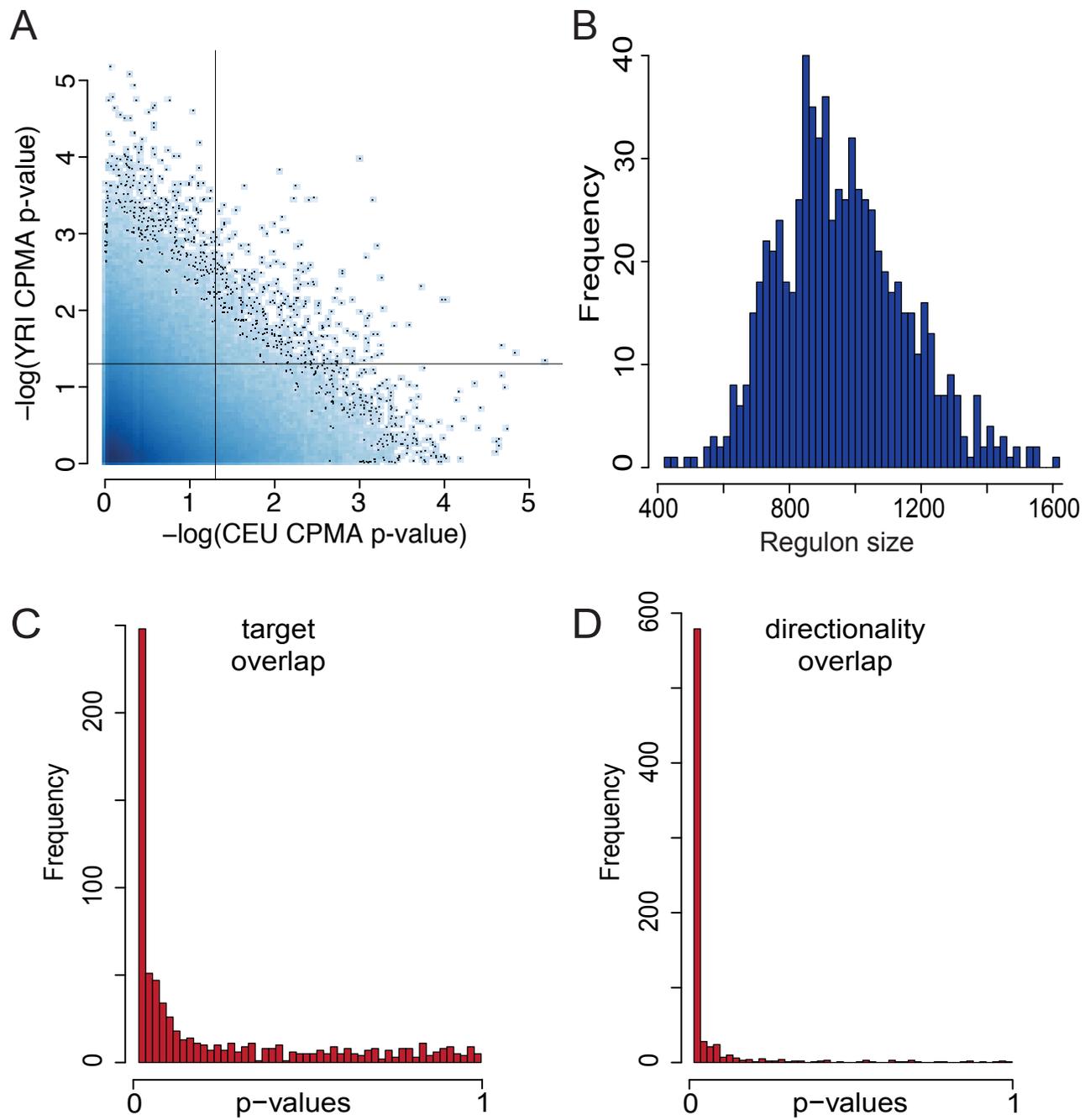

Figure 1

Figure 2

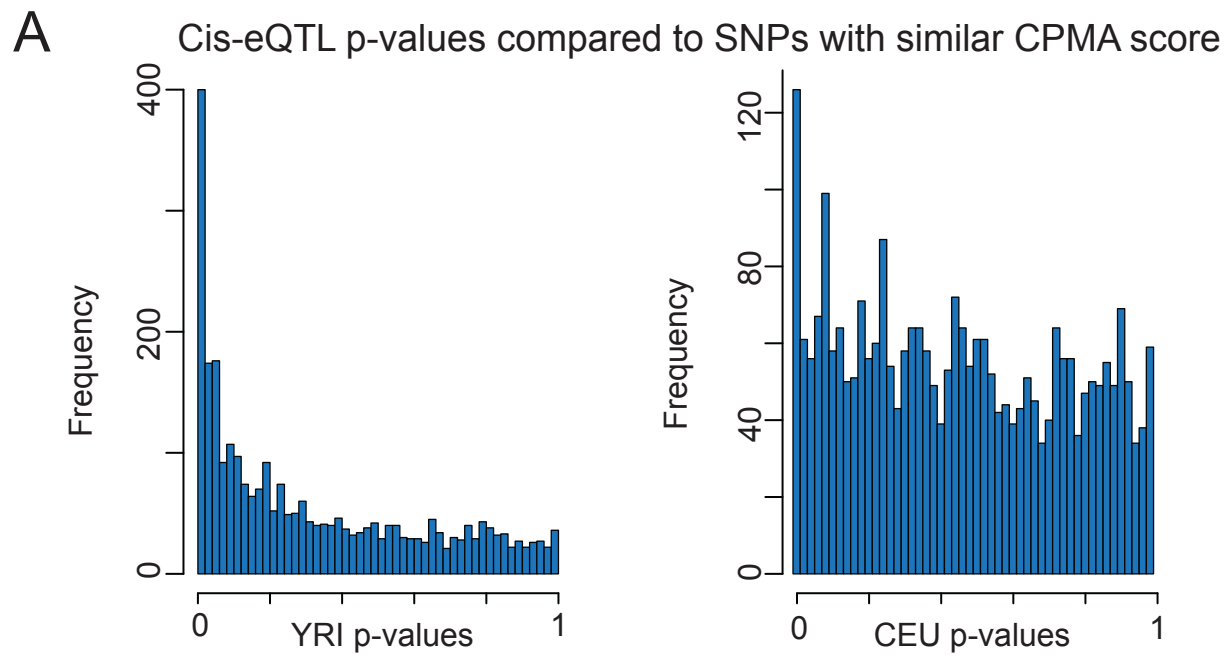

A. Cis-eQTL p-values compared to SNPs with similar CPMA score

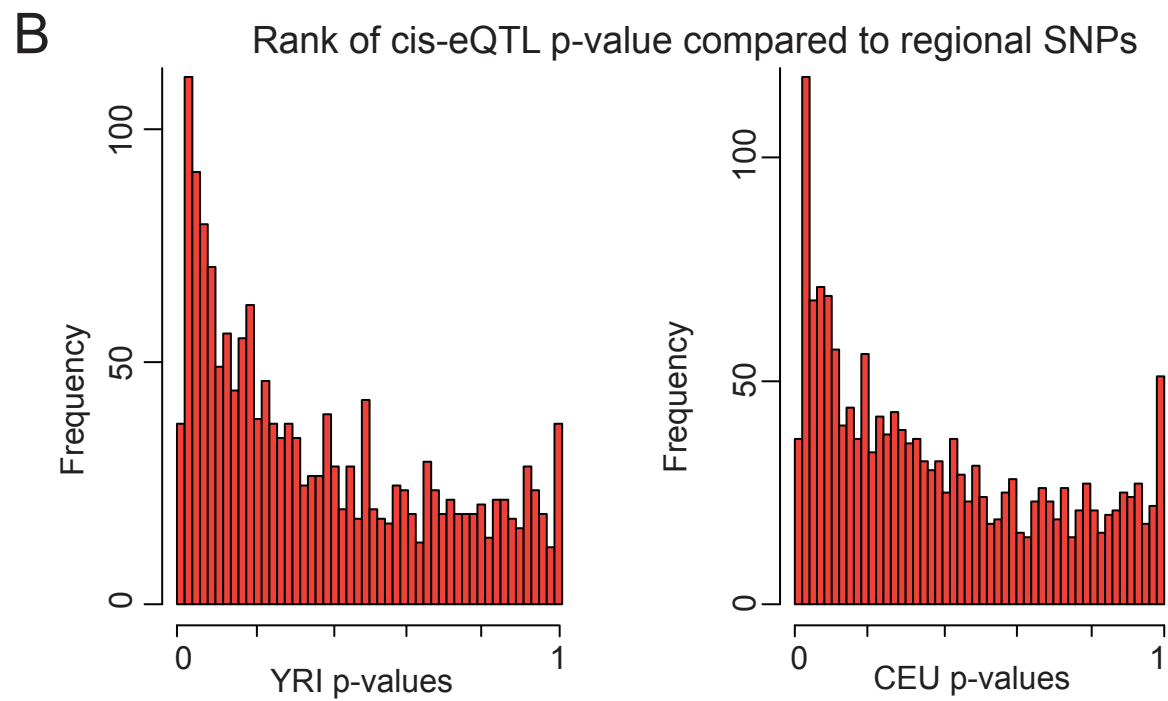

B. Rank of cis-eQTL p-value compared to regional SNPs

Supplementary Figure 1

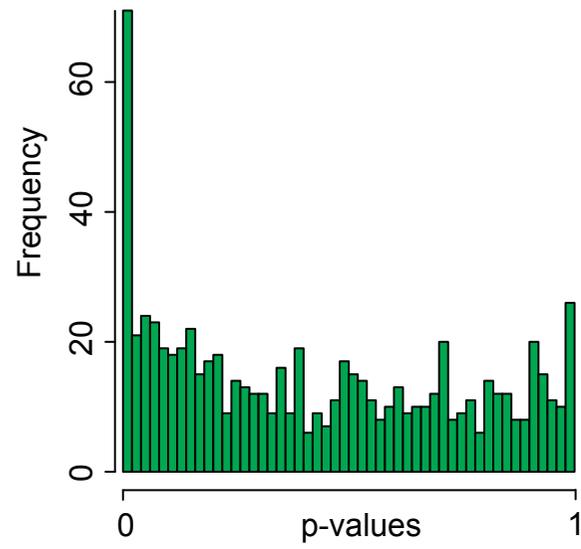 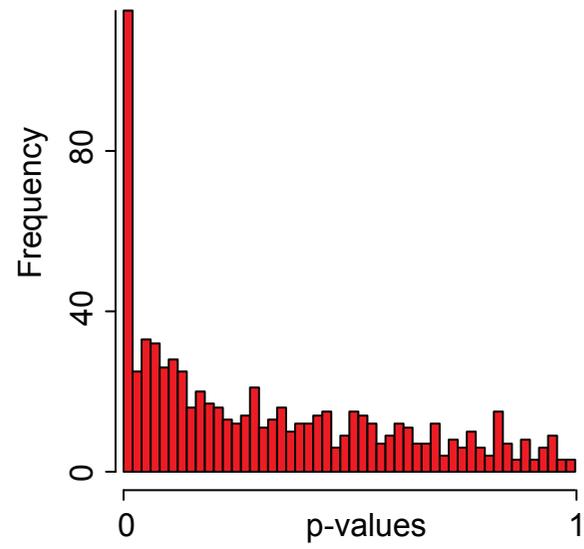

A                                    B

| Variant | | | trans eQTL evidence (CPMA) | | Regulon composition | | | | Effect consistency between populations ($p$) | Target gene connectivity DAPPLE direct edges (p) | Number of TFs enriched | CEU variance explained | | YRI variance explained | |
|---|---|---|---|---|---|---|---|---|---|---|---|---|---|---|---|
| SNP ID | chr | position | CEU $P$ value | YRI $P$ value | CEU | YRI | Overlap | Expected (95% CI) | Overlap (p-value) | | | | Median | (5% - 95%) | Median | (5% - 95%) |
| rs4654727 | 1 | 21214735 | 0.016 | 0.030 | 2720 | 2664 | 1196 | 861 (773 - 977) | 9.21E-04 | >1.0E-5 | 0.01 | 3 | 2.78% | (0.96 - 6.74) | 3.47% | (1.54 - 7.61) |
| rs3906402 | 1 | 60702323 | 0.047 | 0.050 | 2604 | 3096 | 1249 | 958 (869 - 1076) | 0.002 | >1.0E-5 | 0.01 | 9 | 2.50% | (1.08 - 6.16) | 2.62% | (0.91 - 6.39) |
| rs821418 | 1 | 153284423 | 0.042 | 0.043 | 2454 | 2600 | 1019 | 757 (673 - 871) | 0.003 | 1.43E-04 | 0.05 | 5 | 2.89% | (1.31 - 6.52) | 2.64% | (1.01 - 6.46) |
| rs1618879 | 1 | 167497812 | 0.019 | 0.035 | 2606 | 2819 | 992 | 873 (785 - 990) | 0.047 | 3.38E-04 | 0.04 | 1 | 2.86% | (0.83 - 6.95) | 2.74% | (1.33 - 6.25) |
| rs11804999 | 1 | 248322479 | 0.011 | 0.008 | 2641 | 2612 | 986 | 819 (733 - 935) | 0.019 | 5.12E-05 | 0.03 | 2 | 3.29% | (1.56 - 8.37) | 4.13% | (2.18 - 8.86) |
| rs2309821 | 2 | 100900304 | 0.001 | 0.021 | 2603 | 2802 | 1412 | 866 (779 - 983) | 4.09E-05 | >1.0E-5 | 0.02 | 3 | 6.02% | (3.12 - 12.15) | 3.20% | (1.41 - 6.51) |
| rs13427703 | 2 | 234247508 | 0.012 | 0.043 | 2692 | 2570 | 965 | 822 (735 - 937) | 0.029 | 1.23E-04 | 0.01 | 4 | 3.55% | (1.42 - 8.54) | 2.99% | (1.37 - 6.71) |
| rs13426909 | 2 | 235512068 | 0.005 | 0.045 | 2807 | 2746 | 1043 | 916 (827 - 1033) | 0.041 | 2.56E-04 | 0.01 | 2 | 3.50% | (1.38 - 7.87) | 2.67% | (1.26 - 6.04) |
| rs9853343 | 3 | 181504232 | 0.001 | 0.045 | 3017 | 2416 | 1004 | 866 (779 - 981) | 0.032 | 1.33E-04 | 0.01 | 5 | 4.97% | (2.40 - 10.55) | 2.63% | (1.37 - 6.21) |
| rs12507259 | 4 | 5112978 | 0.045 | 0.005 | 2369 | 2666 | 958 | 750 (665 - 863) | 0.008 | 7.16E-05 | 0.02 | 6 | 2.77% | (1.30 - 6.08) | 4.82% | (2.27 - 10.98) |
| rs6534412 | 4 | 124441989 | 0.046 | 0.015 | 2760 | 2905 | 1089 | 953 (863 - 1071) | 0.036 | 1.33E-04 | 0.01 | 1 | 1.96% | (0.44 - 5.12) | 3.05% | (1.36 - 7.53) |
| rs155375 | 5 | 39312722 | 0.016 | 0.025 | 2593 | 2772 | 1178 | 854 (766 - 970) | 0.001 | >1.0E-5 | 0.01 | 6 | 3.02% | (1.17 - 7.45) | 2.97% | (1.52 - 7.47) |
| rs6866215 | 5 | 53906687 | 0.004 | 0.035 | 2667 | 2566 | 1226 | 813 (726 - 928) | 2.25E-04 | >1.0E-5 | 0.01 | 7 | 3.50% | (1.41 - 8.09) | 3.19% | (1.57 - 7.15) |
| rs7736745 | 5 | 117089085 | 0.040 | 0.039 | 2657 | 2872 | 1029 | 907 (818 - 1024) | 0.045 | 1.02E-05 | 0.01 | 2 | 2.61% | (1.11 - 6.09) | 2.75% | (1.21 - 6.49) |
| rs9275596 | 6 | 32681631 | 0.025 | 0.021 | 2785 | 2707 | 1059 | 896 (807 - 1013) | 0.021 | 4.09E-05 | 0.01 | 2 | 2.82% | (1.16 - 6.37) | 3.22% | (1.65 - 6.89) |
| rs2268730 | 6 | 52402823 | 0.029 | 0.036 | 2494 | 2605 | 905 | 771 (686 - 886) | 0.034 | 1.43E-04 | 0.04 | 3 | 2.83% | (0.94 - 7.99) | 3.03% | (1.14 - 7.68) |
| rs1936017 | 6 | 72624622 | 0.021 | 0.045 | 2434 | 2897 | 971 | 837 (751 - 953) | 0.035 | 2.25E-04 | 0.01 | 1 | 3.22% | (1.59 - 6.99) | 2.23% | (0.77 - 5.61) |
| rs13219185 | 6 | 82079170 | 0.041 | 0.003 | 2437 | 2947 | 1043 | 853 (766 - 969) | 0.013 | >1.0E-5 | 0.01 | 2 | 2.40% | (0.51 - 7.13) | 4.31% | (1.97 - 8.77) |
| rs6960872 | 7 | 144702641 | 0.016 | 0.026 | 2580 | 2820 | 1137 | 864 (777 - 981) | 0.003 | 3.07E-05 | 0.04 | 6 | 3.41% | (1.54 - 8.20) | 3.08% | (1.33 - 7.05) |
| rs10810330 | 9 | 1512369 | 0.049 | 0.025 | 2381 | 2212 | 743 | 625 (54 - 732) | 0.040 | >1.0E-5 | 0.01 | 4 | 2.82% | (1.12 - 9.08) | 4.07% | (2.04 - 9.68) |
| rs2200032 | 9 | 32005136 | 0.026 | 0.043 | 2917 | 2855 | 1414 | 990 (900 - 1109) | 2.56E-04 | >1.0E-5 | 0.01 | 2 | 2.54% | (0.81 - 7.13) | 2.99% | (1.17 - 6.71) |
| rs7049166 | 9 | 112381955 | 0.004 | 0.020 | 2780 | 2956 | 1321 | 977 (887 - 1095) | 9.62E-04 | >1.0E-5 | 0.03 | 6 | 4.11% | (1.86 - 8.79) | 3.03% | (1.36 - 6.83) |
| rs2418763 | 10 | 107301067 | 0.001 | 0.038 | 2878 | 2951 | 1441 | 1010 (919 - 1129) | 2.46E-04 | >1.0E-5 | 0.01 | 3 | 5.00% | (2.41 - 9.89) | 2.71% | (1.18 - 5.88) |
| rs999136 | 10 | 121829203 | 0.021 | 0.032 | 2357 | 2685 | 952 | 751 (666 - 864) | 0.009 | >1.0E-5 | 0.01 | 1 | 3.39% | (1.59 - 8.61) | 2.97% | (1.40 - 7.22) |
| rs10766619 | 11 | 20058525 | 0.028 | 0.048 | 2655 | 2844 | 1100 | 897 (809 - 1015) | 0.011 | >1.0E-5 | 0.02 | 5 | 2.97% | (1.27 - 7.16) | 2.36% | (0.92 - 6.14) |
| rs2900511 | 12 | 23011944 | 0.007 | 0.031 | 2606 | 2820 | 1060 | 873 (785 - 990) | 0.014 | 5.12E-05 | 0.02 | 11 | 3.59% | (1.69 - 8.30) | 3.02% | (1.31 - 6.46) |
| rs10878883 | 12 | 69375536 | 9.67E-05 | 0.007 | 2566 | 2932 | 1394 | 894 (806 - 1011) | 8.19E-05 | >1.0E-5 | 0.02 | 6 | 7.44% | (3.98 - 14.08) | 4.16% | (2.04 - 8.40) |
| rs10846902 | 12 | 125981163 | 0.006 | 0.020 | 2571 | 2932 | 1098 | 896 (807 - 1013) | 0.011 | 1.54E-04 | 0.02 | 6 | 4.22% | (2.08 - 9.26) | 3.28% | (1.40 - 7.87) |
| rs7323636 | 13 | 104973086 | 0.036 | 0.015 | 2617 | 2883 | 1063 | 896 (808 - 1014) | 0.020 | 1.23E-04 | 0.01 | 2 | 2.86% | (0.89 - 6.48) | 2.48% | (0.69 - 5.82) |
| rs4356419 | 15 | 50320288 | 0.050 | 0.013 | 2634 | 2865 | 1117 | 897 (808 - 1014) | 0.008 | 8.19E-05 | 0.05 | 4 | 2.51% | (1.08 - 5.88) | 3.52% | (1.76 - 7.25) |
| rs961229 | 15 | 96725043 | 0.015 | 0.003 | 2394 | 3170 | 1231 | 902 (814 - 1018) | 0.001 | 1.02E-05 | 0.02 | 1 | 3.69% | (1.56 - 9.90) | 4.44% | (2.15 - 9.04) |
| rs4572372 | 15 | 100745994 | 0.044 | 0.031 | 2515 | 2794 | 1069 | 835 (748 - 950) | 0.006 | >1.0E-5 | 0.03 | 2 | 2.63% | (1.20 - 6.22) | 2.71% | (1.28 - 7.19) |
| rs9912204 | 17 | 6166020 | 0.025 | 0.016 | 2267 | 2756 | 1029 | 741 (657 - 854) | 0.002 | >1.0E-5 | 0.03 | 5 | 3.76% | (1.68 - 8.98) | 3.42% | (1.57 - 8.54) |
| rs2154611 | 22 | 24989920 | 0.035 | 0.021 | 2503 | 2776 | 949 | 825 (738 - 941) | 0.042 | 4.27E-03 | 0.01 | 2 | 2.88% | (1.16 - 7.38) | 2.88% | (1.39 - 6.91) |
| rs11089447 | 22 | 30794719 | 0.038 | 0.005 | 2554 | 2737 | 1034 | 830 (743 - 946) | 0.010 | 5.12E-05 | 0.02 | 4 | 2.71% | (1.06 - 6.82) | 4.47% | (2.33 - 9.32) |
| rs5980159 | X | 15489506 | 0.004 | 0.036 | 2327 | 2774 | 967 | 766 (681 - 879) | 0.009 | 1.02E-04 | 0.04 | 5 | 3.96% | (1.47 - 8.07) | 3.51% | (1.11 - 7.85) |

| Regulon | | Transcription factor binding enrichment | | Correlation with regulon member expression | |
|---|---|---|---|---|---|
| *trans*-acting SNP | Targets in regulon | TF | *p* value | CEU | YRI |
| rs2900511 | 1060 | SIX5 | 0.009 | 0.52 | 0.60 |
| | | ZNF143 | 0 | 0.28 | ns |
| | | ELF1 | 0.044 | 0.65 | 0.80 |
| | | EBF1 | 0.025 | 0.41 | 0.67 |
| | | MEF2A | 0.033 | 0.58 | 0.32 |
| | | RAD21 | 0.034 | ns | ns |
| | | GABPA | 0.024 | -0.29 | -0.73 |
| | | BCL11A | 0.032 | ns | 0.27 |
| | | ATF3 | 0.03 | 0.60 | 0.76 |
| | | NFKB1 | 0.003 | 0.71 | 0.69 |
| rs2418763 | 1441 | SP1 | 0.004 | ns | 0.37 |
| | | NFKB1 | 0.006 | 0.67 | 0.66 |
| rs10766619 | 1100 | CTCF | 0.029 | 0.35 | ns |
| | | GABPA | 0.021 | -0.25 | -0.65 |
| | | NFKB1 | 0.025 | 0.69 | 0.68 |
| | | NR2C2 | 0.019 | ns | -0.21 |
| rs10846902 | 1098 | SP1 | 0.008 | 0.21 | 0.44 |
| | | SIX5 | 0.025 | 0.47 | 0.57 |
| | | ZNF143 | 0.006 | 0.26 | ns |
| | | GABPA | 0.002 | -0.29 | -0.68 |
| rs10878883 | 1394 | ZNF143 | 0.033 | 0.27 | ns |
| | | EBF1 | 0.048 | 0.39 | 0.65 |
| | | GABPA | 0.02 | -0.28 | -0.71 |
| | | YY1 | 0.007 | 0.58 | 0.44 |
| | | SRF | 0.044 | 0.63 | -0.29 |
| | | NR2C2 | 0.047 | ns | -0.23 |
| rs10810330 | 743 | CTCF | 0.038 | 0.59 | ns |
| | | SIX5 | 0.027 | 0.45 | ns |
| | | ZNF143 | 0.011 | 0.34 | 0.36 |
| | | USF1 | 0.042 | ns | -0.37 |
| rs6960872 | 1137 | SP1 | 0.018 | ns | 0.32 |
| | | BCL3 | 0.027 | 0.53 | 0.63 |
| | | BCLAF1 | 0.029 | ns | -0.49 |
| | | POU2F2 | 0.044 | 0.73 | 0.88 |
| rs2200032 | 1414 | YY1 | 0.032 | 0.51 | 0.42 |
| rs7049166 | 1321 | CTCF | 0.026 | 0.24 | -0.21 |
| | | SP1 | 0.015 | 0.21 | 0.40 |
| | | GABPA | 0 | -0.36 | -0.71 |
| | | USF1 | 0.024 | -0.36 | -0.38 |
| | | USF2 | 0.047 | 0.74 | 0.74 |
| rs5980159 | 967 | SP1 | 0.012 | ns | 0.35 |
| | | POU2F2 | 0.027 | 0.81 | 0.91 |
| | | USF1 | 0.045 | -0.28 | -0.31 |
| rs11089447 | 1034 | SP1 | 0.002 | 0.21 | 0.38 |
| | | GABPA | 0.026 | -0.27 | -0.70 |
| | | YY1 | 0.019 | 0.57 | 0.40 |
| rs2154611 | 949 | MAX | 0.008 | ns | -0.32 |
| rs4356419 | 1117 | SP1 | 0.041 | ns | 0.38 |
| | | GABPA | 0.003 | -0.34 | -0.74 |
| rs7323636 | 1063 | SP1 | 0.028 | ns | 0.37 |
| rs9912204 | 1029 | SIX5 | 0.036 | 0.53 | 0.46 |
| | | MAX | 0.004 | ns | -0.23 |
| | | FAM48A | 0.021 | -0.31 | ns |
| | | NR2C2 | 0 | -0.30 | -0.30 |
| | | ZZZ3 | 0.041 | ns | ns |
| rs961229 | 1231 | GABPA | 0.024 | -0.48 | -0.75 |
| rs13426909 | 1043 | GABPA | 0.007 | -0.34 | -0.73 |
| | | YY1 | 0.026 | 0.56 | 0.42 |
| rs13427703 | 965 | BRCA1 | 0.016 | -0.39 | -0.29 |
| rs12507259 | 958 | SP1 | 0.016 | ns | 0.33 |
| | | EP300 | 0.03 | 0.75 | 0.28 |
| | | YY1 | 0.036 | 0.56 | 0.31 |
| | | TCF12 | 0.036 | -0.41 | ns |
| | | IRF4 | 0.024 | -0.26 | ns |
| rs9853343 | 1004 | SP1 | 0.005 | 0.25 | 0.44 |
| | | ELF1 | 0.03 | 0.67 | 0.82 |
| | | GABPA | 0.001 | -0.26 | -0.68 |
| | | USF1 | 0.044 | -0.30 | -0.39 |
| rs1618879 | 992 | BCL3 | 0.012 | 0.47 | 0.51 |
| rs821418 | 1019 | SP1 | 0.039 | ns | 0.36 |
| | | GABPA | 0.005 | -0.21 | -0.67 |
| | | BCL3 | 0.029 | 0.54 | 0.62 |
| rs4654727 | 1196 | SP1 | 0.016 | ns | 0.34 |
| rs3906402 | 1249 | SP1 | 0 | ns | 0.39 |
| | | SIX5 | 0.05 | 0.49 | 0.64 |
| | | EP300 | 0.048 | 0.77 | 0.35 |
| | | GABPA | 0.018 | -0.22 | -0.69 |
| | | BCL11A | 0.042 | ns | 0.30 |
| | | BCL3 | 0.042 | 0.52 | 0.62 |
| | | POU2F2 | 0.005 | 0.83 | 0.88 |
| rs2309821 | 1412 | GABPA | 0.037 | -0.45 | -0.76 |
| rs11804999 | 986 | MEF2A | 0.006 | 0.64 | 0.33 |
| rs2268730 | 905 | USF1 | 0.034 | -0.19 | -0.20 |
| | | IRF3 | 0.025 | ns | ns |
| rs9275596 | 1059 | GABPA | 0.01 | -0.36 | -0.73 |
| | | BCLAF1 | 0.046 | ns | -0.55 |
| rs7736745 | 1029 | GABPA | 0.003 | -0.25 | -0.65 |
| rs1936017 | 971 | SRF | 0.039 | 0.67 | -0.26 |
| rs6866215 | 1226 | SP1 | 0.002 | ns | 0.32 |
| | | MEF2A | 0.039 | 0.60 | 0.31 |
| | | BCL3 | 0.036 | 0.49 | 0.59 |
| | | BRCA1 | 0.039 | -0.23 | ns |
| rs155375 | 1178 | GABPA | 0 | -0.34 | -0.74 |
| | | BCL11A | 0.008 | ns | 0.26 |
| | | BCLAF1 | 0.047 | ns | -0.48 |
| | | TCF12 | 0.03 | -0.40 | 0.23 |
| | | NFKB1 | 0.047 | 0.63 | 0.60 |

| trans-eQTL | chr | position | Potential cis regulator | Evidence |
|---|---|---|---|---|
| rs4654727 | 1 | 21214736 | HP1BP3 | part of the regulon |
|  |  |  | EIF4G3 | cis-correlation significant |
| rs821418 | 1 | 153284424 | SPRR1A | cis-correlation significant |
|  |  |  | GATAD2B | cis-correlation significant |
| rs11804999 | 1 | 248322480 | OR2T35 | cis-correlation significant |
| rs13427703 | 2 | 234247509 | DGKD | cis-correlation significant |
| rs12507259 | 4 | 5112979 | MSX1 | cis-correlation significant |
| rs9275596 | 6 | 32681632 | HLA-DRB5 | part of the regulon, *cis*-eQTL |
|  |  |  | HLA-DRB1 | part of the regulon, *cis*-eQTL |
|  |  |  | PSMB9 | part of the regulon |
| rs2268730 | 6 | 52402824 | MCM3 | part of the regulon, cis-correlation significant, significant DAPPLE seed |
|  |  |  | RN7SK | cis-correlation significant |
|  |  |  | ICK | cis-correlation significant |
| rs999136 | 10 | 121829204 | SEC23IP | part of the regulon, *cis*-eQTL |
| rs10766619 | 11 | 20058526 | PRMT3 | part of the regulon |
| rs2900511 | 12 | 23011945 | ETNK1 | part of the regulon |
| rs10878883 | 12 | 69375537 | NUP107 | part of the regulon, *cis*-eQTL |
| rs4356419 | 15 | 50320289 | USP8 | cis-correlation significant |
| rs4572372 | 15 | 100745995 | MEF2A | cis-correlation significant |
| rs11089447 | 22 | 30794720 | MTMR3 | part of the regulon |
|  |  |  | PES1 | part of the regulon |
|  |  |  | SF3A1 | cis-correlation significant |
|  |  |  | DUSP18 | cis-correlation significant |
| rs2154611 | 22 | 24989921 | SPECC1L | cis-correlation significant |
|  |  |  | ADORA2A | cis-correlation significant |
|  |  |  | UPB1 | cis-correlation significant |